\documentclass[copyright,creativecommons]{eptcs}
\providecommand{\event}{GandALF 2010} 
\usepackage{breakurl}             

\usepackage{listings}
\usepackage{array}
\usepackage{multicol}
\usepackage{multirow}
\usepackage{amsmath}
\usepackage{graphicx}
\usepackage[latin1]{inputenc}
\newcolumntype{V}[1]{>{\small \raggedright}p{#1}}
\newcolumntype{N}{>{\small}l}
\newtheorem{definition}{Definition}
\newtheorem{theorem}{Theorem}

\title{Efficient Symmetry Reduction and the Use of State Symmetries for Symbolic Model Checking}
\author{Christian Appold
\institute{Chair of Computer Science V\\
University of Würzburg\\
Würzburg, Germany}
\email{appold@informatik.uni-wuerzburg.de}
}
\def\titlerunning{Efficient Symmetry Reduction and the Use of State Symmetries for Symbolic Model Checking}
\def\authorrunning{Christian Appold}
\begin{document}
\maketitle

\begin{abstract}
One technique to reduce the state-space explosion problem in temporal logic model checking is symmetry reduction. The combination of symmetry reduction and symbolic model checking by using BDDs suffered a long time from the prohibitively large BDD for the orbit relation. Dynamic symmetry reduction calculates representatives of equivalence classes of states dynamically and thus avoids the construction of the orbit relation. In this paper, we present a new efficient model checking algorithm based on dynamic symmetry reduction. Our experiments show that the algorithm is very fast and allows the verification of larger systems. We additionally implemented the use of state symmetries for symbolic symmetry reduction. To our knowledge we are the first who investigated state symmetries in combination with BDD based symbolic model checking.
\end{abstract}

\section{Introduction}
\label{sec:Introduction}
With the growing dispersion of concurrent systems, e.g. through the use of multi-core CPUs or sensor networks, the need for reliable methods for their verification increases. A successful technique for the verification of concurrent systems which exhaustively examines the state-space of a system is temporal logic model checking \cite{TempModCheckEmCl}, \cite{TempModCheckQuSif}. Model checking is an automated formal verification technique, where properties are formulated in a temporal logic (like CTL \cite{BranchingTimeLogic} or LTL \cite{LTLPnueli}).  Unfortunately model checking suffers from the state-space explosion problem. This especially appears in the verification of concurrent systems. There, the size of the state-space grows exponentially with the number of components. Concurrent systems often contain many replicated components (e.g. sensor networks often consist of hundreds of nodes). But they frequently also possess a lot of symmetries. Symmetry reduction techniques \cite{IpDill96BetterVerif} have been developed to exploit those symmetries and to combat the state-space explosion problem. In many cases significant savings in memory and time can be achieved by using them (see e.g. \cite{Ip93efficientverification}). 

Symmetry reduction techniques exploit symmetries by restricting state-space search to representatives of equivalence classes of states.  One key problem of symmetry reduction in model checking is to calculate that states are in the same equivalence class. This problem is known as the orbit problem. The authors of \cite{ExploitSymmetryClarke96} have proven that it is at least as hard as the graph isomorphism problem, which is very difficult to solve. With the help of the orbit relation, model checking can be done with a bisimilar quotient structure over the equivalence classes (see e.g. \cite{ExploitSymmetryClarke96}, \cite{Emerson94symmetry}). Symmetry reduction has been first introduced in explicit-state model checking. An explicit-state model checker that uses symmetry reduction is for example Murphi \cite{IpDill96BetterVerif}. In symbolic model checking with BDDs, which has been very successful in the verification of large systems, exploiting symmetry becomes more complex. The reason therefore is that the orbit relation has to be represented as a BDD. The size of this BDD is exponential in the minimum of the number of components and the number of states per component for many frequently occurring symmetry groups \cite{ExploitSymmetryClarke96}. Consequently symbolic model checking with symmetry reduction and a BDD for the orbit relation can be used only for systems with a small number of components or where each component has only a few states.

One method which avoids to build the orbit relation is to use multiple representatives for each orbit \cite{ExploitSymmetryClarke96}. Although the multiple representatives approach has been an improvement, it is still not good enough to verify systems of interesting size. The reason is that when using multiple representatives the state-space of the quotient model is not reduced as much as with unique representatives. Additionally the BDD which relates states to their representatives is generally still very large. Another technique for symbolic model checking of fully symmetric systems by using BDDs is to use generic representatives \cite{EmersonTrefler99fromasymmetry}. Therewith the orbit problem and the construction of the orbit relation can be avoided. In this approach the original program text is translated into a reduced program, which can be explored with standard model checking algorithms without further symmetry considerations. A global state of the reduced program is a vector of counters, with one counter for each local state. The counter indicates the number of processes which are currently in this state. The approach is more generally known as counter abstraction \cite{PnuelliXuZuck02LivenessCounterAbstraction}. The authors of \cite{EW03SymmetryGeneric} extended the approach to include systems with global shared variables. If generic representatives are applicable, their usage is very effective and compares well  to the unique or multiple representatives approach. However they suffer from the local state-space explosion problem, and the translation to a counter abstracted program can be difficult, too. 

A technique where orbit representatives are calculated dynamically during fixpoint iterations is dynamic symmetry reduction \cite{EmersonWahlDynSymRed05}. There transition images are computed with respect to the unreduced structure and successor states are immediately mapped afterwards to the corresponding orbit representatives. Dynamic symmetry reduction is not restricted to fully symmetric systems and can handle data and component symmetry. Experimental results have shown that the approach often outperforms the use of multiple and generic representatives. Another advantage to the unique or multiple representatives approach is that only representatives for states which actually occur during the state-space traversal have to be generated and stored. The performance bottleneck of this technique is the swapping of bits in the BDD representation of the model, which is necessary for a representative calculation.

Dynamic symmetry reduction as presented in \cite{EmersonWahlDynSymRed05} uses a single BDD for the transition relation. This BDD contains all transitions of every component of the input program. In \cite{BurchClarkeLong91SymbModPartTransRel} the authors showed how a partitioned transition relation can be used instead. Therewith they have been able to verify systems which could not be verified by using a single unpartitioned transition relation, because it would be intractably large. In verification experiments, where verification also failed with a partitioned transition relation, larger portions of the state-space could be investigated.

In this paper we propose a new efficient symbolic model checking algorithm for forward reachability analysis, which uses dynamic symmetry reduction. As suggested by \cite{BurchClarkeLong91SymbModPartTransRel}, to achieve the verifiability of larger systems, our algorithm does not use only a single transition relation. Instead, we always store simultaneously only the transition relation of one component of a concurrent system. Therewith our algorithm is able to verify systems where the whole transition relation cannot be build due to memory exhaustion. This is especially useful in combination with symmetry reduction, which enables the verification of systems with many replicated components. With our algorithm we extend their usability for systems with a larger number of replicated components and also a huge single transition relation.  Through the combination of component-wise execution and full exploration of new states for one component before the execution of the next component we achieve considerable runtime improvements for dynamic symmetry reduction. Also the component-wise execution of transitions helps to implement state symmetries efficiently. State symmetries use the internal symmetries of a single global state to avoid redundant calculations of orbit representatives. They have been first introduced in \cite{Emerson94symmetry}. The authors of \cite{SistlaEmersonFair97} integrated their use into an explicit-state model checking algorithm. Especially in the verification of systems with many replicated components big runtime savings can be achieved by using them. As far as we know, we are the first which investigated state symmetries in symbolic model checking with BDDs. For our verification experiments we used and extended the symbolic model checker Sviss \cite{WahlBlancEmerson08Sviss}, which implements symbolic symmetry reduction methods. As our experimental results show (see section \ref{subsec:ExpResults}), state symmetries can often considerably improve the runtime of our symbolic model checking algorithm. 

The rest of the paper is organized as follows. In the next section we present some background information that is useful throughout the paper. We there give an introduction to model checking (\ref{subsec:modelchecking}), symmetry reduction (\ref{sec:SymmetryReduction}), dynamic symmetry reduction (\ref{sec:DynamicSymmetryReduction}) and state symmetries (\ref{subsec:stateSymmetry}). In Section \ref{subsec:FastSymRed} we present our new fast model checking algorithm for dynamic symmetry reduction, before we describe our implementation of state symmetries in Section \ref{subsec:ImplStateSymmetries}. Experimental results which confirm the efficiency of our algorithm and the usefulness of state symmetries are presented in Section \ref{subsec:ExpResults}. The paper closes with a conclusion and an outlook to future work.

\section{Background}
\label{subsec:Background}

\subsection{Model Checking}
\label{subsec:modelchecking}
Model checking \cite{TempModCheckEmCl}, \cite{TempModCheckQuSif} is an automatic technique to verify finite state concurrent systems. Given a finite state model describing the behavior of a  system and a property, a model checker determines if the property is satisfied by the model.  The finite state model of a system is usually described in the form of a \textit{Kripke structure}.  
\begin{definition}
Let AP be a finite set of atomic propositions. A Kripke  structure M over AP is a quadruple 
\newline
$M=(S,R,L,S_{0})$, with the following components:
\begin{itemize}
\item $S$ is a nonempty, finite set of states,
\item $R \subseteq S \times S$ is the transition relation, 
\item $L: S \rightarrow 2^{AP}$ is a function, which maps each state in $S$ with the set of atomic propositions which are true in that state and
\item $S_{0} \subseteq S$ is the set of initial states.
\end{itemize}
\end{definition}
Properties are usually specified in a temporal logic. Examples of temporal logics are CTL and LTL, which are sublogics of the temporal logic CTL* \cite{CTLSternEH}. They extend propositional logic with temporal operators.  

\subsection{Symmetry Reduction}
\label{sec:SymmetryReduction}
This section gives an introduction to symmetries in model checking. For further information see e.g. \cite{SymmetryMillDon06} or \cite{ExploitSymmetryClarke96}. A Kripke structure is symmetric if it is invariant under certain transformations of its state-space. Permutations are used to define symmetries of a Kripke structure. Given a non-empty set $X$, a permutation of $X$ is a bijection $\pi \, : \, X \rightarrow X$. We extend $\pi$ to a mapping $\pi \, : R \rightarrow R$ on the transition level of a Kripke structure by defining $\pi((s,t)) = (\pi(s),\pi(t))$.
\newpage
\begin{definition}
A permutation $\pi$ on $S$ is said to be a \textbf{symmetry} of a Kripke structure $M=(S,R,L,S_{0})$, if:
\begin{itemize}
\item $R$ is invariant under $\pi$ : $\pi(R)=R$, 
\item $L$ is invariant under $\pi \, : \, L(s) = L(\pi(s))$ for any $s \in S$, and
\item $S_{0}$ is invariant under $\pi$ : $\pi(S_{0}) = S_{0}$.
\end{itemize}
The symmetries of $M$ form a group under function composition. A model $M$ is said to be \textbf{symmetric}, if its symmetry group $G$ is non-trivial (i.e. does not consist only of the identity permutation). 
\end{definition}
In a concurrent system with $n$ replicated components a $state \, (\vec g, l_{1},..., l_{n})$ consists of the values $\vec g$ of all global variables (not associated with any process) and the $local \, state \, l_{i}$ of each process $i \in \{1,...,n\}$ (values of all local variables of process $i$). There are different types of symmetries. Common ones are component symmetry and data symmetry. In component symmetry a symmetry $\pi$ is derived from a permutation on $\{1,...,n\}$ and acts on a state $s=(\vec g, l_{1},..., l_{n})$ as $\pi (s)=(\vec g^{\pi}, l_{\pi(1)},..., l_{\pi(n)})$. The local states of the processes are permuted by permuting their positions in the state vector. Further, $\pi$ acts on $\vec g$ by acting component-wise on each global variable $g$. The action of $\pi$ on $g$ depends on the nature of $g$, for more details see \cite{EW03SymmetryGeneric}. Under data symmetry \cite{IpDill96BetterVerif} $\pi$ acts on data values, in the form $\pi(\vec g, l_{1},..., l_{n})=(\pi(\vec g), \pi{(l_{1})},..., \pi{(l_{n})})$. As an example for the difference between component symmetry and data symmetry consider the permutation $\pi$ on $\{a,b\}$, which exchanges $a$ and $b$. For the state $(a,a)$ the application of component symmetry by exchanging positions $1$ and $2$ of the state leads to the same state $(a,a)$.  With data symmetry we get the state $(b,b)$ through application of $\pi$, which exchanges the values of $a$ and $b$. 

 A group $G$ of symmetries induces an equivalence relation $\equiv_{G}$ on the states of $M$ by the rule $s \equiv_{G} t \Leftrightarrow s = \pi(t)$ for some $\pi \in G$. The equivalence class of a state $s \in S$ under $\equiv_{G}$, denoted $[s]_{G}$, is called the $orbit$ of $s$ under the action of $G$.  The relation $\equiv_{G}$  is called orbit relation. Observe that $s \equiv_{G} t$ implies $L(s) = L(t)$, since $L$ is invariant under permutations of $G$ (see Definition 2). The orbits can be used to construct a $quotient$ Kripke structure $M_{G}$. 
\begin{definition}
The quotient Kripke structure $M_{G}$ of $M$ with respect to $G$ is a quadruple $M_{G} = (S_{G},R_{G},L_{G},S^{0}_{G})$ where:  
\begin{itemize}
\item $S_{G}=\{[s]_{G} \, : \, s \in S\}$ (the set of orbits of $S$ under the action of $G$),
\item $R_{G}=\{([s]_{G},[t]_{G}) \, : \, (s,t) \in R\}$ (quotient transition relation),
\item $L_{G}([s]_{G})=L(rep_{G}([s]_{G}))$ (where $rep_{G}([s]_{G})$ is a unique representative of $[s]_{G}$),
\item $S^{0}_{G}= \{[s]_{G} \, : \, s \in S_{0}\}$ (the orbits of the initial states $S_{0}$ under the action of $G$).
\end{itemize}
\end{definition}
In practice for the set $S_{G}$ the set of orbit representatives is taken instead of the orbits themselves. The quotient structure $M_{G}$ is smaller than $M$, if $G$ is non-trivial. For any $s$, the size of $[s]_{G}$ is bounded by $|G|$, so the theoretical minimum size of $S_{G}$ is $|S| / |G|$. In highly symmetric systems we may have $|G|=n!$, where $n$ is the number of components. It has been shown that $M$ and $M_{G}$ are equivalent in the sense that they satisfy the same set of logic properties which are invariant under permutations of $G$.
A proof of the following theorem can be found in \cite{ExploitSymmetryClarke96}.
\begin{theorem}
Let $M=(S,R,L,S_{0})$  be a Kripke Structure, $G$ be a symmetry group of $M$, and $h$ be a CTL* formula. If $h$ is invariant under the group $G$, then
\begin{equation}
M,s \models h  \Leftrightarrow M_{G},[s]_{G}  \models h
\end{equation}
where $M_{G}$ is the quotient structure corresponding to $M$.
\label{symLab}
\end{theorem}
As a consequence, by choosing a suitable symmetry group $G$, model checking can be done by using $M_{G}$ instead of $M$, which often leads to considerable savings in memory and time (see e.g. \cite{ExploitSymmetryClarke96}).

\subsection{Dynamic Symmetry Reduction}
\label{sec:DynamicSymmetryReduction}
In this subsection we explain dynamic symmetry reduction (for more information see \cite{EmersonWahlDynSymRed05}) for symbolic forward reachability analysis with BDDs. Dynamic symmetry reduction calculates orbit representatives dynamically during state-space traversal. Therewith the computation of the orbit relation which often is of intractable size can be avoided. Also only representatives which actually occur during state-space traversal (which might be few) have to be maintained, in contrast to the unique or multiple representatives approach. Advantages to the generic representatives approach are that dynamic symmetry reduction is not restricted to fully symmetric systems and also no possibly complicated transformations of the input program are necessary.
Listing \ref{lst:fixAbst} shows the standard forward reachability analysis fixpoint routine supplemented with dynamic symmetry reduction. For comparison, the fixpoint routine which uses the quotient transition relation can be seen in Listing \ref{lst:fixOld}. The orbit relation thereby is essentially embedded in the BDD for the quotient transition relation. Therefore their computation, even if the orbit relation is not used directly for it, is in a reasonable amount of time in general only possible for very simple verification examples.

  \begin{minipage}[b]{5cm}
 \begin{lstlisting}[language=C++,numbers=left,showspaces=false,showtabs=false,showstringspaces=false,mathescape=true,captionpos=b,caption=Fixpoint routine for forward reachability analysis with quotient transition relation,label=lst:fixOld]
Y = Init; 
do {
  Y' = Y;
  Y = Init $\lor$ $Image_{\,R_{G}\,}$Y; }
$\textbf{while}$(Y != Y');
return Y;
\end{lstlisting}
  \end{minipage}
\hspace{3cm}
  \begin{minipage}[b]{5cm}

 \begin{lstlisting}[captionpos=left,language=C++,numbers=left,showspaces=false,showtabs=false,showstringspaces=false,mathescape=true,captionpos=b,caption=Fixpoint routine for forward reachability analysis with dynamic symmetry reduction,label=lst:fixAbst]
Z = Init; 
do {
  Z' = Z;
  Z = Init $\lor$ $\alpha$($Image_{\,R\,}$Z); }
$\textbf{while}$(Z != Z');
return Z;
\end{lstlisting}
  \end{minipage}

 In Listing \ref{lst:fixAbst} an operator $\alpha$ is used instead of the expensive quotient transition relation in Listing \ref{lst:fixOld}. The operator $\alpha$ is applied to the result of the forward image operation $Image_{\,R\,}Z$ with the unreduced transition relation $R$. It is an abstraction operator which dynamically maps states that result from the forward image computation to their corresponding representatives. Equation \ref{eq:alpha} shows the formal definition of $\alpha$. Depending on the underlying group of symmetry, the implementation of the abstraction function $\alpha$ has to be adapted. 

\begin{equation} \label{eq:alpha} \alpha(T) = \{rep_{G}([t]_{G}) \in S_{G} : \exists t \in T : (t,rep_{G}([t]_{G}) ) \: \in \: \equiv_{G} \}  \end{equation}

In the following we describe the underlying algorithm of $\alpha$ for the most common and most important case of full component symmetry. Systems often have id-sensitive global variables whose values are component ids.  An example therefore is a component which currently has an exclusive copy of some cache data. Under full component symmetry usually the lexicographically least element of an orbit is chosen as representative for the orbit. This element can be found through sorting of the local state vector of a given global state. If global id-sensitive variables are available, also a rule to get representative values for them is required. For example consider a system with three components and one global id-sensitive variable and the two states (A,B,B,2) and (A,B,B,3). The global id-sensitive variable is listed last here. As one can see the vector with the local states of the processes is already lexicographically sorted. But if we would not consider the special role of global id-sensitive variables, we would have no unique representative. By choosing the maximum value for them, (A,B,B,3) is the unique representative.

In dynamic symmetry reduction where BDDs are used, in contrast to explicit-state model checking, not only a single global state, but a set of global states has to be sorted simultaneously. If a global state is not  sorted correctly, dynamic symmetry reduction swaps the BDD variable order to gain the unique representative of the state. The authors of \cite{EmersonWahlDynSymRed05} say that swapping of bits in the BDD representation dominates efficiency of dynamic symmetry reduction. The complexity of the swap operations thereby depends exponentially on the distance between the bits in the BDD variable ordering that have to be swapped. Therefore they propose to use bubble sort which swaps only adjacent elements.

  \begin{minipage}[b]{0.35 \linewidth}
 \begin{lstlisting}[language=C++,numbers=left,showspaces=false,showtabs=false,showstringspaces=false,mathescape=true,captionpos=b,caption= $\alpha(T)$ for unique representatives computation with dynamic symmetry reduction under full component symmetry,label=lst:fixSort]
Z = T; 
do {
  Z' = Z;
  Z = $\tau(Z)$; }
$\textbf{while}$(Z != Z');
return Z;
\end{lstlisting}
  \end{minipage}
\hspace{2cm}
  \begin{minipage}[b]{0.35 \linewidth}
 \begin{lstlisting}[language=C++,numbers=left,showspaces=false,showtabs=false,showstringspaces=false,mathescape=true,captionpos=b,caption= $\tau(Z)$ for unique representatives computation with dynamic symmetry reduction under full component symmetry,label=lst:fSortSwap]
for(p=1;p<=(n-1);p++) { 
  $Z_{bad}$ = Z $\land$ $\neg \{z:p \leq_{z} p+1\}$;
  if($Z_{bad}$ $\neq$ $\emptyset$) {
    $Z_{good}$ = Z \ $Z_{bad}$;
    $Z_{swapped}$ = swap(p,p+1,$Z_{bad}$);
    Z = $Z_{good}$ $\lor$ $Z_{swapped}$; }
}
return Z;
\end{lstlisting}
  \end{minipage}

The corresponding symbolic sorting algorithm can be seen in Listing \ref{lst:fixSort} and Listing \ref{lst:fSortSwap}. There $p \leq_{z} p+1$ means for a state $z$ and the local states of components $p$ and $p+1$ that either $l_{p}(z) < l_{p+1}(z)$, or $l_{p}(z) = l_{p+1}(z)$ and none of the id-sensitive global variables has value $p$. This rule is for representatives with global id-sensitive variables with maximum values and $l_{i}(z)$ is the local state of component $i$. For $n$ components and $\leq_{z}$ the set of representatives is then defined as:

\begin{equation} \label{eq:repStates} rep_{G}(S)=\{z \in S : \forall p < n : p \leq_{z} p + 1\} = \bigcap_{p < n} \{z \in S : p \leq_{z} p + 1\}. \end{equation}

The algorithm of Listing \ref{lst:fixSort} is executed, when the abstraction function $\alpha(T)$ is applied. It iteratively executes the algorithm of Listing \ref{lst:fSortSwap} until a fixpoint is reached. Then the unique representatives of the orbits of the states from $T$ have been calculated. Each time the algorithm of Listing \ref{lst:fSortSwap} is called, it looks for states where the components are not in correct order with respect to $\leq_{z}$ and executes the necessary BDD swaps. Thereby it stores states where $\leq_{z}$ is violated in $Z_{bad}$. If $Z_{bad}$ is not empty, the necessary swaps in the BDD variable ordering are done for all states in $Z_{bad}$ simultaneously.  As mentioned before, the swapping of bits in the BDD variable ordering is also the expensive step of the algorithm. Beneath further information an extension of the dynamic symmetry reduction principle to full CTL model checking can be found in \cite{EmersonWahlDynSymRed05}.
 
\subsection{State Symmetries}
\label{subsec:stateSymmetry}
Beneath symmetry reduction state symmetries (see \cite{SistlaEmersonFair97} and \cite{SMC99}) can also be used in model checking of concurrent systems with replicated components. They use the internal symmetries of a single global  state of the Kripke structure.  Up to now state symmetries have only been investigated for explicit-state model checking. 

A model checker which uses state symmetries in explicit-state model checking is for example SMC \cite{SMC99}. It is able to use symmetry reduction and also state symmetries. In SMC two components $i$ and $j$ are said to be \textit{equivalent in a global state s}, if $\theta_{i,j}(s)=s$, where $\theta_{i,j}$ is the permutation that interchanges $i$ and $j$ but fixes all other components. This relation is an equivalence relation among the components, which induces a partition on the set of components and is called the \textit{state symmetry partition of s}. All components in the same local state and to which the described permutation can be applied and leads to $\theta_{i,j}(s)=s$ are in the same group of this partition. If SMC uses state symmetries it only executes every enabled transition for one representative component of every group of a \textit{state symmetry partition of s}. It does not execute executable transitions from other components of the partition group,  because the same transitions are executable for each component of a group and the representatives of their successor states would be the same. Thus therewith the overhead of redundant representative calculations can be avoided. Due to their restrictive notion of state symmetries they have been able to use efficient algorithms for their detection. The authors of \cite{SMC99} also presented some experimental results. Especially when systems with a large number of components have been verified, they observed significant runtime improvements through state symmetries.

The authors of \cite{Appold09UsingStateSym} investigated the use of state symmetries for the explicit-state model checker Murphi \cite{IpDill96BetterVerif} and presented some enhancements to the use of state symmetries. They also could achieve considerable runtime savings in their experiments by using them. A less restrictive notion of state symmetries has been proposed by \cite{Emerson94symmetry}, but they did not present experimental results. 

\section{Our new fast Model Checking Algorithm}
\label{subsec:FastSymRed}
In this section we present our new fast symbolic forward reachability analysis algorithm for dynamic symmetry reduction. The pseudo-code of the algorithm can be found in Listing \ref{lst:Fastsym}. First of all, the BDD  named \textit{Init} is initialized with the initial states of the verification model. The initial states are immediately sorted (line 2) by using the abstraction function $\alpha$ of dynamic symmetry reduction (see section \ref{sec:DynamicSymmetryReduction}). By sorting of the initial states we achieve that the first forward image computation explores only successors of symmetry reduced states, even if the given initial states were unsorted. This circumvents redundant swaps of bits in the BDD representation for successors of not symmetry reduced initial states which would not lead to new unique representatives. Next, one BDD for successor states during forward image computation (\textit{Successors}) and one BDD that later saves all states which have been reached during the state-space traversal (\textit{Reached}) are generated and both initialized with the initial states (see line 3).  Then one BDD for the transition relation (\textit{TransRelation}) and another BDD that stores states which have been reached during the current exploration of a component (\textit{newExplored}) are generated (see line 4). The array of BDDs \textit{toExplore} in line 6 stores for each component of each component type the states which have still to be explored for this component. The value of \textit{compTypes} thereby is the number of different component types in the verification model and \textit{maxCompNumber} is the maximum number of components that appears for a component type.  The array \textit{compNum[]} (see e.g. line 9) contains for each component type the number of available components. At the beginning, \textit{toExplore} is for every component initialized with the sorted initial states.

In line 14 a loop starts and will be executed until there is no component that has any further states to explore. In the loop the transition relation for the currently active component is build on-the-fly (line 17). An advantage of our algorithm is that always only the transition relation of the currently active component has to be stored. This is the component for which states are explored at the moment through forward image calculations. BDDs for the transition relation of the other components are build not until they are needed. This saves a lot of memory, especially if the transition relation is large. Therewith, we can even verify systems which cannot be verified by using a single transition relation, because a single transition relation would be too large to be build (see e.g. the peterson mutual exclusion protocol in section \ref{subsec:ExpResults}). 
 
\begin{minipage}[b]{15cm}
\begin{multicols}{2}
\begin{lstlisting}[language=C++,numbers=left,showspaces=false,showtabs=false,showstringspaces=false,captionpos=b,caption=Pseudo-code of our fast forward reachability analysis algorithm,label=lst:Fastsym,basicstyle=\scriptsize,mathescape]
BDD Init = initialStates();
Init = $\alpha$(Init);
BDD Successors$,$ Reached = Init;
BDD TransRelation$,$ newExplored = Empty();
bool finish = false; 
BDD toExplore[compTypes][maxCompNumber];

for(i=0;i<compTypes;i++) { 
  for(j=(compNum[i]-1);j>=0;j$\text{\--\--}$) { 
    toExplore[i][j] = Init;	
  }
}

while(finish == false) {
 for(i=0;i<compTypes;i++) { 
  for(j=(compNum[i]-1);j>=0;j$\text{\--\--}$) { 
      TransRelation = buildTransRel(i,j);       
      Successors = toExplore[i][j];
      
      while(Successors != Empty()) {
        Successors = $Image_{R}$(Successors);
        Successors = $\alpha$(Successors) 
            $\&$ !newExplored $\&$ !Reached;
        newExplored |= Successors;
      }

      Reached |= newExplored;
     


      for(z=0;z<compTypes;z++) { 
        for(k=(compNum[z]-1);k>=0;k$\text{\--\--}$) { 
          if(k != j || z != i) {
            toExplore[z][k] |= newExplored; 
          }	
          else {
            toExplore[z][k] = Empty(); 
          }
        }
      }
      newExplored = Empty();
	  
       $\textbf{/* here we integrated}$ 
         $\textbf{the use of state symmetries (see Listing \ref{lst:sSymm}) */}$

    }
  }

  finish = true;
  for(n=0;n<compTypes;n++) { 
    for(m=(compNum[n]-1);m>=0;m$\text{\--\--}$) { 
      if(toExplore[n][m] != Empty()) {
        finish = false; }
    }
  }
}
\end{lstlisting}
\end{multicols}
\end{minipage}

In line 20 a loop begins which is executed as long as new states can be found for the currently active component. Inside the loop forward images ($Image_R(Successors)$) are calculated with states that have not been explored for the component so far. Afterwards unique representatives of the successor states are computed (line 22). Representatives which have not been visited during state-space traversal are saved in the BDD \textit{Successors} and further explored for the component. 

The multiple consecutive application of the forward image computation for one component has the advantage that in this way successor states often can be canonicalized considerably faster. In dynamic symmetry reduction exploration of states always starts from symmetry reduced states. By execution of transitions for only one component, less changes of these symmetry reduced states occur than by using the whole transition relation with all components for forward image computation. Therefore fewer swaps are needed to canonicalize these successor states, which reduces the time for their canonicalization. Also, all newly found states for one component are added to \textit{toExplore} of the other components after full exploration of the component. Therewith \textit{toExplore} can contain a large amount of states if the component which executes transitions changes. Necessary BDD swaps then can be used for a larger amount of states simultaneously. Together, as our experimental results confirm (see section \ref{subsec:ExpResults}), considerable runtime improvements can be achieved.

In line 32 and 35 the discovered new states are added to \textit{toExplore} of the other components of the system. Whenever all components have explored their states (the loop in line 15 has finished), the algorithm tests, if there still is a component which has to explore some states. If no such component can be found, all states which are reachable from the initial states have been found and the algorithm terminates. The correctness of the algorithm follows from the fact, that every newly discovered global state is added first to \textit{newExplored} and after the full exploration of a component to \textit{toExplore} of all other components. Therewith forward images of this state are calculated for the component which discovered this global state and for all other components.

\section{Implementation of State Symmetries}
\label{subsec:ImplStateSymmetries}
In explicit-state model checking state symmetries can lead to large runtime improvements, especially when systems with many replicated components are verified.  We implemented the use of state symmetries for our new algorithm and the case of fully symmetric systems in the model checker Sviss \cite{WahlBlancEmerson08Sviss}. For other symmetry groups, e.g. rotational symmetry, state symmetries can also be used. However, the computation can possibly be more complex sometimes and therefore more time consuming. In the worst case even an increase in runtime could appear.

\begin{center}
\begin{minipage}[b]{10cm}
\begin{lstlisting}[language=C++,numbers=left,showspaces=false,showtabs=false,showstringspaces=false,captionpos=b,caption=Our implementation of state symmetries,label=lst:sSymm,basicstyle=\scriptsize,mathescape]
BDD stateSymmStates=Empty();

//j is the index of the currently active component
if(j!=0) {
  $\textbf{for\_each}$(global_idst_Var) {
      stateSymmStates = stateSymmStates | equal(global_idst_Var$,$j);
      stateSymmStates = stateSymmStates | equal(global_idst_Var$,$j-1);
   }
}

stateSymmStates = toExplore[i][j-1] & !stateSymmStates;
stateSymmStates = stateSymmStates & equal(j,j-1);

toExplore[i][j-1] = toExplore[i][j-1] & !stateSymmStates;
\end{lstlisting}
\end{minipage}
\end{center}

The pseudo-code of our implementation of state symmetries can be found in Listing \ref{lst:sSymm}. In our experiments we started the component-wise exploration for each component type with the component with the largest component index. For this component no state symmetries have to be computed. The reason is that even in the presence of state symmetries there has to be a component which executes all enabled transitions of a state symmetry group. In our implementation we have chosen for it the component with the highest component index.

To gain the most runtime benefits from state symmetries, it is important to detect them fast. Also it is advantageous to detect additionally as much state symmetries as possible. The experimental results we present in section \ref{subsec:ExpResults} have been achieved by inserting state symmetry detection at line 43 of Listing \ref{lst:Fastsym}, after the full exploration of a component. We always calculated state symmetries between two neighboring components. For example if currently component $i+1$ has been explored fully, we detected state symmetries in \textit{toExplore} of component $i$ and there between the components $i$ and $i+1$. This can be done efficiently and is possible because the local states of components are sorted lexicographically in symmetry reduced global states of fully symmetric systems. Therefore in general the local state bits of components with possible state symmetries are neighbors in a symmetry reduced global state. If we detected any state symmetries between the neighboring components, we removed the corresponding global states from \textit{toExplore} of $i$ (line 14). Consequently, in the presence of state symmetries always the component with the higher component index explores new states for a global state with state symmetries. During its forward image calculations component $i$ then has not to consider successors of states with state symmetries to component $i+1$. Therewith we target to avoid superfluous BDD swaps. Calculation of successors of such states for component $i$ would only lead to already visited representatives. It is worth to mention that superfluous BDD swaps are only avoided at all in one canonicalization step of dynamic symmetry reduction, if component $i$ has not explored other global states whose canonicalization needs these swaps during the current forward image calculation. If the canonicalization of another global state requires BDD swaps which could be avoided for a global state by the use of state symmetries, the swaps cannot be saved and have to be executed for this global state. But our experimental results show in spite of this peculiarity of state symmetries in symbolic model checking, in contrast to explicit-state model checking, also significant runtime improvements can be achieved.

We also made some experiments where state symmetries have always been calculated for every component and not only between neighbors. This has the advantage that therewith all redundant canonicalizations due to existing state symmetries could be eliminated. In the approach presented before, global states where state symmetries exist between two components can sometimes be explored for both components. This occurs for example if a component discovers a new global state where a state symmetry exists between this component and another component. The component then explores this global state immediately and the global state is also added to \textit{toExplore} of the other component. If this component already has explored its global states in this turn of the algorithm, it also explores this state in the next turn. Our experimental results showed  that with this state symmetry implementation nearly no runtime gains could be achieved for our new algorithm. In contrast the implementation strategy mentioned in the last paragraph often leads to large runtime improvements. Therewith to gain the biggest runtime improvements in symbolic model checking, it is also necessary to choose an efficient implementation of state symmetries.

In our implementation we used state symmetries in the presence of global id-sensitive variables only, if no such variable pointed to one of the neighboring components (see lines 6,7 and 11 in Listing \ref{lst:sSymm}). Therewith also some state symmetries could be lost depending on the verification model, but computation of state symmetries could be much more complex if such cases would be considered, too. This again could diminish possible runtime gains. As mentioned in section \ref{subsec:FastSymRed} in our new model checking algorithm all new states which have been discovered for a component are added to \textit{toExplore} of all other components. With the help of states symmetries we achieve that states with state symmetries can be deleted from \textit{toExplore} before their exploration. Thus state symmetries help our algorithm to avoid redundant swaps of bits in the BDD variable ordering.

\section{Experimental Results}
\label{subsec:ExpResults}
Here we present the results of our verification experiments. We have done the experiments on a computer with an Intel Pentium Core 2 CPU with 2.4 GHz and 3 GB main memory by using a single core.  As operating system we used Debian 4.0. The verification experiments have been done with the symbolic model checker Sviss, which uses the Cudd BDD package \cite{Somenzi09Cudd}. For our experiments we disabled dynamic variable reordering of BDDs. As variable order for the bits of the components in the BDDs we have chosen the variable order concatenated:
\begin{gather*} \label{eq:lexOrder2} 
\mbox{\textbf{concatenated:} $b_{1 1}$ ... $b_{1 log l}$ $b_{2 1}$ ... $b_{2 log l}$ ...... $b_{n 1}$ ... $b_{n log l}$}
\end{gather*}
Here $b_{i j}$ denotes the $j$th bit of component $i$ and $l$ is the number of local states of a component. In the following tables the number of components in the verification experiments can be found in the column \textit{Problem} after the name of the verification benchmark. \textit{Number of BDD Nodes} is the  largest number of live BDD nodes that appeared during a verification experiment. This is the memory bottleneck of a verification experiment, because the model checker has to store this number of BDD nodes to finish verification successfully. Time is the runtime of a verification experiment, where s, m and h are abbreviations for seconds, minutes and hours.

\begin{figure}[htb]
	\centering
	\includegraphics[width=5.5cm]{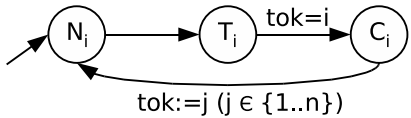}
	\caption{Synchronization skeleton of the simple mutual exclusion example}
	\label{lst:syncSkel}
\end{figure}

Table \ref{table:MutEx} presents verification results for verification experiments with a simple mutual exclusion example. We have chosen this testcase because it allows to test both our new algorithm and the use of state symmetries in symbolic model checking for a very large number of components. A synchronization skeleton \cite{TempModCheckEmCl} of the testcase can be found in Figure \ref{lst:syncSkel}. Every component has only the three local states \textit{non-critical ($N$)}, \textit{trying ($T$)} and \textit{critical ($C$)}. There are state changes from \textit{non-critical} to \textit{trying} and from \textit{critical} to \textit{non-critical} which can be executed without restrictions. Also there is a global id-sensitive variable \textit{tok}, which ranges over process indices. 
Its value is set nondeterministically to a process index, if a component executes a state change from \textit{critical} to \textit{non-critical}. Only the process whose id currently is the value of the global id-sensitive variable is allowed to make a state change from \textit{trying} to \textit{critical}. For our verification experiments we used the property that no two processes are in the state \textit{critical} simultaneously. The verification results show that large runtime and memory improvements can be achieved by using our new model checking algorithm. The use of state symmetries lead to further runtime reductions. Particularly in verification experiments with a large number of components big runtime improvements could be observed for this testcase.

\begin{table}[htbp]
\centering
\begin{tabular}{|c|c|c|c|c|c|c|} \hline
\multicolumn{1}{|V{3cm}|}{} & \multicolumn{2}{V{3cm}|}{Old Dynamic Symmetry Reduction} & \multicolumn{2}{V{3cm}|}{New Algorithm Only} & \multicolumn{2}{V{3cm}|}{New Algorithm With State Symmetries} \\ \hline
\multicolumn{1}{|V{3cm}|}{Problem} & \multicolumn{1}{V{1.8cm}|}{Number of BDD Nodes} & \multicolumn{1}{V{1.2cm}|}{Time} & \multicolumn{1}{V{1.8cm}|}{Number of BDD Nodes} & \multicolumn{1}{V{1.2cm}|}{Time} & \multicolumn{1}{V{1.8cm}|}{Number of BDD Nodes} & \multicolumn{1}{V{1.2cm}|}{Time}\\ \hline
\multicolumn{1}{|V{3cm}|}{Mutex 200} & \multicolumn{1}{r|}{337,709} & \multicolumn{1}{r|}{4:25m} & \multicolumn{1}{r|}{54,684} & \multicolumn{1}{r|}{22s} & \multicolumn{1}{r|}{54,684} & \multicolumn{1}{r|}{14s}\\ \hline
\multicolumn{1}{|V{3cm}|}{Mutex 400} & \multicolumn{1}{r|}{2,044,109} & \multicolumn{1}{r|}{56:32m}  & \multicolumn{1}{r|}{189,702} & \multicolumn{1}{r|}{3:53m} & \multicolumn{1}{r|}{189,702} & \multicolumn{1}{r|}{1:50m}\\ \hline
\multicolumn{1}{|V{3cm}|}{Mutex 600} & \multicolumn{1}{r|}{2,932,995} & \multicolumn{1}{r|}{4:39h} & \multicolumn{1}{r|}{405,133} & \multicolumn{1}{r|}{13:34m} & \multicolumn{1}{r|}{405,133} & \multicolumn{1}{r|}{6:26m}\\ \hline
\multicolumn{1}{|V{3cm}|}{Mutex 800} & \multicolumn{1}{r|}{5,190,561} & \multicolumn{1}{r|}{14:10h} & \multicolumn{1}{r|}{700,120} & \multicolumn{1}{r|}{33:35m} & \multicolumn{1}{r|}{700,120} & \multicolumn{1}{r|}{17:11m}\\ \hline
\end{tabular}
\caption{Verification results for the simple mutual exclusion example}
\label{table:MutEx}
\end{table}

In Table \ref{table:CCPMCS} experimental results of verification experiments with MCSLock, a modified variant of the list-based queuing algorithm from \cite{MellorScott91AlgForScalSync}, can be found. In this example the number of local states of a process is small and we used for our experiments the property that no two processes can possess the lock at the same time. The experimental results show significant runtime improvements by using our algorithm. Also additional runtime gains through state symmetries could be observed. The runtime could be even more than halved and reduced very much for systems with a large number of components, as the experiment with 60 components shows. Table \ref{table:CCPMCS} also shows experimental results for the CCP cache coherence protocol. It refers to a cache coherence protocol developed from S. German (see for example \cite{PnueliRuahZuck01AutomDeducVer}). This protocol is characterized by components with a large number of local states. Nevertheless our algorithm is nearly twice as fast as the previous dynamic symmetry reduction algorithm. Also the memory requirements could be reduced significantly. State symmetries there do not lead to similarly large runtime gains as before. The reason possibly is that BDD swaps, which can be saved through the use of state symmetries, cannot be saved at all, because they are needed to sort other global states.

\begin{table}[htbp]
\centering
\begin{tabular}{|c|c|c|c|c|c|c|} \hline
\multicolumn{1}{|V{3cm}|}{} & \multicolumn{2}{V{3cm}|}{Old Dynamic Symmetry Reduction} & \multicolumn{2}{V{3cm}|}{New Algorithm Only} & \multicolumn{2}{V{3cm}|}{New Algorithm With State Symmetries} \\ \hline
\multicolumn{1}{|V{3cm}|}{Problem} & \multicolumn{1}{V{1.8cm}|}{Number of BDD Nodes} & \multicolumn{1}{V{1.2cm}|}{Time} & \multicolumn{1}{V{1.8cm}|}{Number of BDD Nodes} & \multicolumn{1}{V{1.2cm}|}{Time} & \multicolumn{1}{V{1.8cm}|}{Number of BDD Nodes} & \multicolumn{1}{V{1.2cm}|}{Time}\\ \hline
\multicolumn{1}{|V{3cm}|}{MCSLock 10} & \multicolumn{1}{r|}{24,251} & \multicolumn{1}{r|}{3s} & \multicolumn{1}{r|}{9,333} & \multicolumn{1}{r|}{2s} & \multicolumn{1}{r|}{8,870} & \multicolumn{1}{r|}{1s}\\ \hline
\multicolumn{1}{|V{3cm}|}{MCSLock 20} & \multicolumn{1}{r|}{143,715} & \multicolumn{1}{r|}{2:43m} & \multicolumn{1}{r|}{55,013} & \multicolumn{1}{r|}{1:20m} & \multicolumn{1}{r|}{51,145} & \multicolumn{1}{r|}{55s}\\ \hline
\multicolumn{1}{|V{3cm}|}{MCSLock 40} & \multicolumn{1}{r|}{786,310} & \multicolumn{1}{r|}{1:41h} & \multicolumn{1}{r|}{446,849} & \multicolumn{1}{r|}{1:04h} & \multicolumn{1}{r|}{426,068} & \multicolumn{1}{r|}{33:05m}\\ \hline
\multicolumn{1}{|V{3cm}|}{MCSLock 60} & \multicolumn{1}{r|}{2,087,657} & \multicolumn{1}{r|}{15:56h} & \multicolumn{1}{r|}{1,744,207} & \multicolumn{1}{r|}{12:00h} & \multicolumn{1}{r|}{1,693,866} & \multicolumn{1}{r|}{5:28h}\\ \hline
\multicolumn{1}{|V{3cm}|}{CCP 10}& \multicolumn{1}{r|}{358,127} & \multicolumn{1}{r|}{2:49m} & \multicolumn{1}{r|}{69,462} & \multicolumn{1}{r|}{52s} & \multicolumn{1}{r|}{73,898} & \multicolumn{1}{r|}{51s}\\ \hline
\multicolumn{1}{|V{3cm}|}{CCP 20} & \multicolumn{1}{r|}{2,429,642} & \multicolumn{1}{r|}{1:41h} & \multicolumn{1}{r|}{355,394} & \multicolumn{1}{r|}{43:37m} & \multicolumn{1}{r|}{366,758} & \multicolumn{1}{r|}{43:32m}\\ \hline
\multicolumn{1}{|V{3cm}|}{CCP 25} & \multicolumn{1}{r|}{4,424,644} & \multicolumn{1}{r|}{5:35h} & \multicolumn{1}{r|}{651,706} & \multicolumn{1}{r|}{2:47h} & \multicolumn{1}{r|}{666,619} & \multicolumn{1}{r|}{2:45h}\\ \hline
\multicolumn{1}{|V{3cm}|}{CCP 30} & \multicolumn{1}{r|}{7,220,011} & \multicolumn{1}{r|}{14:36h} & \multicolumn{1}{r|}{1,100,968} & \multicolumn{1}{r|}{8:36h} & \multicolumn{1}{r|}{1,119,322} & \multicolumn{1}{r|}{8:33h}\\ \hline
\end{tabular}
\caption{Verification results for the MCSLock and the CCP example}
\label{table:CCPMCS}
\end{table}

In Table \ref{table:PetRW} experimental results for the peterson mutual exclusion protocol \cite{Peterson81} can be found. In this protocol entry to the critical section is gained by a single process via a series of $n-1$ competitions. There is at least one looser for each competition and the protocol satisfies the mutual exclusion condition, since at most one process can win the final competition. In contrast to the benchmarks before, this protocol has more global id-sensitive variables and also one component has many local states. By using the old dynamic symmetry reduction algorithm verification experiments finished only for a maximum of six components. The reason therefore has been the huge BDD of the single transition relation. It could have been build only for six components. Due to the component-wise treatment of the transition relation in our new model checking algorithm, we could verify the protocol for twelve components. Additionally we achieved large runtime and memory gains for six components. This shows that our new algorithm, beneath its performance advantages, also allows the verification of larger systems. State symmetries here also delivered additional runtime improvements. 

\begin{table}[htbp]
\centering
\begin{tabular}{|c|c|c|c|c|c|c|} \hline
\multicolumn{1}{|V{3cm}|}{} & \multicolumn{2}{V{3cm}|}{Old Dynamic Symmetry Reduction} & \multicolumn{2}{V{3cm}|}{New Algorithm Only} & \multicolumn{2}{V{3cm}|}{New Algorithm With State Symmetries} \\ \hline
\multicolumn{1}{|V{3cm}|}{Problem} & \multicolumn{1}{V{1.8cm}|}{Number of BDD Nodes} & \multicolumn{1}{V{1.2cm}|}{Time} & \multicolumn{1}{V{1.8cm}|}{Number of BDD Nodes} & \multicolumn{1}{V{1.2cm}|}{Time} & \multicolumn{1}{V{1.8cm}|}{Number of BDD Nodes} & \multicolumn{1}{V{1.2cm}|}{Time}\\ \hline
\multicolumn{1}{|V{3cm}|}{Peterson 6} & \multicolumn{1}{r|}{81,931,144} & \multicolumn{1}{r|}{3:22m} & \multicolumn{1}{r|}{186,246} & \multicolumn{1}{r|}{22s} & \multicolumn{1}{r|}{186,246} & \multicolumn{1}{r|}{21s}\\ \hline
\multicolumn{1}{|V{3cm}|}{Peterson 8} & \multicolumn{1}{c|}{mem ov} & \multicolumn{1}{c|}{-} & \multicolumn{1}{r|}{2,385,546} & \multicolumn{1}{r|}{10:24m} & \multicolumn{1}{r|}{2,385,546} & \multicolumn{1}{r|}{10:05m}\\ \hline
\multicolumn{1}{|V{3cm}|}{Peterson 10} & \multicolumn{1}{c|}{mem ov} & \multicolumn{1}{c|}{-} & \multicolumn{1}{r|}{12,810,763} & \multicolumn{1}{r|}{1:40h} & \multicolumn{1}{r|}{12,505,489} & \multicolumn{1}{r|}{1:36h}\\ \hline
\multicolumn{1}{|V{3cm}|}{Peterson 12} & \multicolumn{1}{c|}{mem ov} & \multicolumn{1}{c|}{-} & \multicolumn{1}{r|}{66,938,967} & \multicolumn{1}{r|}{13:22h} & \multicolumn{1}{r|}{65,887,889} & \multicolumn{1}{r|}{12:54h}\\ \hline
\multicolumn{1}{|V{3cm}|}{Readers-Writers 40} & \multicolumn{1}{r|}{65,139} & \multicolumn{1}{r|}{3:30m} & \multicolumn{1}{r|}{18,586} & \multicolumn{1}{r|}{5s} & \multicolumn{1}{r|}{18,586} & \multicolumn{1}{r|}{2s}\\ \hline
\multicolumn{1}{|V{3cm}|}{Readers-Writers 100} & \multicolumn{1}{r|}{393,939} & \multicolumn{1}{r|}{7:53h} & \multicolumn{1}{r|}{100,546} & \multicolumn{1}{r|}{1:41m} & \multicolumn{1}{r|}{100,546} & \multicolumn{1}{r|}{43s}\\ \hline
\multicolumn{1}{|V{3cm}|}{Readers-Writers 200} & \multicolumn{1}{r|}{$>$1,500,000} & \multicolumn{1}{r|}{$>$24h} & \multicolumn{1}{r|}{381,146} & \multicolumn{1}{r|}{21:25m} & \multicolumn{1}{r|}{381,146} & \multicolumn{1}{r|}{6:40m}\\ \hline
\multicolumn{1}{|V{3cm}|}{Readers-Writers 400} & \multicolumn{1}{r|}{$>$2,000,000} & \multicolumn{1}{r|}{$>$36h} & \multicolumn{1}{r|}{1,482,346} & \multicolumn{1}{r|}{4:54h} & \multicolumn{1}{r|}{1,482,346} & \multicolumn{1}{r|}{1:14h}\\ \hline
\end{tabular}
\caption{Verification results for the peterson mutual exclusion protocol and the readers-writers problem}
\label{table:PetRW}
\end{table}

To test the performance of our algorithm on an example with two different component types, we also made experiments with the readers-writers problem (see Table \ref{table:PetRW}). There are multiple readers and writers which share a common memory. For this testcase the number of components means that this number of readers and  also this number of writers has been used in the corresponding verification experiment. In this testcase multiple readers can get access to the shared memory at the same time. If a writer has access to the shared memory, no reader and no other writer should have access to the shared memory. This has also been the property which we used for our verification experiments. In the testcase every reader has only the three local states \textit{idle}, \textit{trying} and \textit{reading}, while every writer has the local states \textit{idle}, \textit{trying} and \textit{writing}. The readers and writers can always execute transitions from \textit{idle} to \textit{trying} and from \textit{reading} and \textit{writing} respectively to \textit{idle}. Readers can execute the transition from \textit{trying} to \textit{reading}, if currently no writer is in the state \textit{writing}. Writers can change their state from \textit{trying} to \textit{writing}, if no reader is in the state \textit{reading} and no other writer is in the state \textit{writing}. Our experimental results show that our model checking algorithm can deliver very big runtime improvements in systems with two components. Further big runtime improvements could be achieved through state symmetries. 

\section{Conclusion and Outlook}
\label{subsec:ConcOutl}
In this paper we propose a new efficient symbolic forward reachability analysis algorithm that allows the efficient use of dynamic symmetry reduction. Through component-wise storing of the transition relation, we achieve the verification of systems where the use of a single transition relation has been intractably large before. Therewith we widened the applicability of dynamic symmetry reduction. Also we presented an approach to integrate the use of state symmetries in our new symbolic model checking algorithm. 

Our experimental results confirm that the new model checking algorithm is considerably faster for all testcases than the usage of dynamic symmetry reduction as presented by \cite{EmersonWahlDynSymRed05}. Additionally our algorithm reduces the memory requirements.  Also the use of state symmetries in symbolic model checking with BDDs can lead to further runtime improvements as investigated before only for explicit-state model checking. 

In the future we will try to find an efficient scheme for component-wise handling of the transition relation and dynamic symmetry reduction for full CTL model checking. At the moment abstraction functions for dynamic symmetry reduction only exist for full symmetry and rotational symmetry. There we want to further enhance the applicability of dynamic symmetry reduction and to test its performance and the performance of our new algorithm for other symmetry groups. Also we plan to investigate methods for the efficient use of symbolic symmetry reduction on multi-core CPUs.

\textit{Related Work}: The closest work to ours is \cite{EmersonWahlDynSymRed05}. There dynamic symmetry reduction has been first presented. An overview about dynamic symmetry reduction has been given in subsection \ref{sec:DynamicSymmetryReduction}. There exists a lot of further work about symmetry reduction for symbolic model checking with BDDs (e.g. multiple representatives and counter abstraction) and the use of state symmetries in explicit-state model checking. More information about these techniques and references for it can be found in section \ref{sec:Introduction} and section \ref{subsec:Background} of this paper. There has already been some work about partitioning the BDD for the transition relation to enable the verification of systems with an otherwise huge BDD for a single transition relation (see e.g. \cite{BurchClarkeLong91SymbModPartTransRel},\cite{MoonKukulaRaviSomen00ToSplitOrToConjoin}).



\bibliographystyle{eptcs} 

\bibliography{Literatur}

\end{document}